\documentclass[a4paper]{spie_neu}  
\usepackage[T1]{fontenc}          
\usepackage{lmodern}                
\usepackage[]{graphicx}
\usepackage{amsmath}
\usepackage{siunitx}
\usepackage{subfigure}
\usepackage[font=small,margin=0.6cm,justification=justified,indention=0.7cm]{caption}[2008/04/01]

\title{Single-photon emission\\from Ni-related color centers\\in CVD diamond}

\author{David Steinmetz*\supit{a}, Elke Neu\supit{a}, Christian Hepp\supit{a}, Roland Albrecht\supit{a}, Wolfgang Bolse\supit{b},\\Jan Meijer\supit{c}, Detlef Rogalla\supit{c} and Christoph Becher\supit{a}
\skiplinehalf
\supit{a}Universit\"at des Saarlandes, Fachrichtung 7.3 (Technische Physik),\\Campus E2.6, 66123 Saarbr\"ucken, Germany;\\
\supit{b}Universit\"at Stuttgart, Institut f\"ur Halbleiteroptik und Funktionelle Grenzfl\"achen,\\Allmandring 3, 70569 Stuttgart, Germany;\\
\supit{c}RUBION, Ruhr-Universit\"at Bochum, Universit\"atsstra\ss e 150, 44780 Bochum, Germany}

\authorinfo{*Send correspondence to d.steinmetz@physik.uni-saarland.de}

\begin{document}

   \maketitle

\begin{abstract}
   Color centers in diamond are very promising candidates among the possible realizations for practical single-photon sources because of their long-time stable emission at room temperature. The popular nitrogen-vacancy center shows single-photon emission, but within a large, phonon-broadened spectrum ($\approx\SI{100}{nm}$), which strongly limits its applicability for quantum communication. By contrast,
   Ni-related centers exhibit narrow emission lines at room temperature. We present investigations on single color centers consisting of Ni and Si created by ion implantation into single crystalline IIa diamond. We use systematic variations of ion doses between $\SI{E8}{cm^{-2}}$ and $\SI{E14}{cm^{-2}}$ and energies between $\SI{30}{keV}$ and $\SI{1,8}{MeV}$. The Ni-related centers show emission in the near infrared spectral range ($\approx\SI{770}{nm}\text{ to }\SI{787}{nm}$) with a small line-width ($\approx\SI{3}{nm}$ FWHM). A measurement of the intensity correlation function proves single-photon emission. Saturation measurements yield a rather high saturation count rate of $\SI{77.9}{kcounts/s}$. Polarization dependent measurements indicate the presence of two orthogonal dipoles.
\end{abstract}



\section{INTRODUCTION}

In recent years, single-photon sources have attracted great interest because of the numerous possible applications in quantum information processing, including quantum metrology, quantum computation and quantum communication. As an example the \emph{BB84}-protocol for quantum key distribution\cite{Bennett1984} (QKD) is based on the usage of single quantum systems to guarantee absolute security against an eavesdropper. If single photons are used as these ``quantum systems'', one requires a single-photon source that can be triggered, features a high emission rate to assure fast information transmission and exhibits a low spectral line-width to allow an effective signal filtering. If weak laser pulses are used as quasi-single photons a general problem arises due to their Poissonian photon statistics, i.\,e. that one pulse can contain more than one photon even if their mean photon number is smaller than one. In principle, the most obvious way to overcome this is the use of true single-photon sources, that can be realized by single emitting quantum systems. If such a dipole undergoes cycles of pulsed excitation, light emission and re-excitation before emitting the next photon, triggered single-photon emission can be realized. Admittedly, many of the existing demonstrations like single ions\cite{Keller2004}, single atoms\cite{Kuhn2002}, single molecules\cite{Brunel1999}, single quantum dots\cite{Santori2001} or even carbon nanotubes\cite{Hoegele2008} hold experimental complexity that limits their suitability for practical applications.

Compared to these candidates, color centers in diamond have the greatest potential for feasible single-photon sources. Color centers are point defects, that consist of combinations of impurity atoms and vacancies in the carbon lattice of diamond. These complexes can form discrete electronic levels within the band gap, which has a size of \SI{5.5}{eV} and makes diamond transparent in the visible and infrared spectrum\cite{Zaitsev2001}. Because of its superior hardness, diamond exhibits a high Debye temperature of \SI{1860}{K}\cite{Burns2009}, so there is only small coupling of the color centers to the surrounding lattice leading to an atom-like spectrum consisting of narrow emission even at high temperatures\cite{Zaitsev2001}. In addition the high photostability of many color centers offers the great advantage of long-time stable single-photon emission at room temperature (RT)\cite{Lounis2005}.

Among the over 500 known color centers in diamond, there are three candidates for single-photon emitters that are intensively explored. The so called \emph{Nitrogen-Vacancy} center (NV) is the most investigated one (and used in the first commercial single-photon source\footnote{Quantum Communications Victoria: http://qcvictoria.com/}). Beyond single-photon emission\cite{Kurtsiefer2000} several key experiments towards quantum information have been performed yet, including the realization of a two-qubit gate \cite{Jelezko2004a} and multipartite entanglement\cite{Neumann2008}. However, NV centers have the detrimental property of a broad emission bandwidth of $\approx\SI{100}{nm}$ due to vibronic coupling to the diamond lattice -- the zero-phonon line only amounts to \SI{3}{\%} of the whole emission rate. This fact strongly limits their suitability for applications in quantum information. The \emph{Silicon-Vacancy} center (SiV) shows enhanced spectral properties ($\approx\SI{10}{nm}$ line-width) but a long-lived shelving state leads to a decreased maximum emission rate\cite{Wang2006} which does not make the SiV very promising, either.

A very auspicious kind of color centers for single-photon sources are nickel-related centers. \emph{NE8} centers that consist of one Ni atom and four N atoms were first found in HPHT diamonds where Ni is used as a catalyst for the crystal growth\cite{Nadolinny1999}. Single Ni-related color centers have been observed in natural IIa type diamond, showing a \SI{1.2}{nm} narrow line around \SI{800}{nm} at RT\cite{Gaebel2004}, but have also been fabricated by CVD growth of a poly-crystalline diamond film on a fused-silica substrate seeded with a slurry containing Ni\cite{Rabeau2005}. For single Ni-related emitters, photon anti-bunching\cite{Gaebel2004} and triggered single-photon emission\cite{Wu2007} have been demonstrated. Investigations of color centers consisting of Ni and possibly Si in nano-crystals yielded high saturation count rates of \SI{200}{kcounts/s} at an emission wavelength of \SI{768}{nm} with an excited-state lifetime of \SI{2}{ns}\cite{Aharonovich2009}.

However, targeted placing of single emitters for applications in quantum information, e.\,g. for coupling to photonic nanostructures\cite{Kreuzer2008}, requires their deterministic creation in bulk diamonds. We present the results of experiments on Ni-related color centers which are created via ion implantation in single crystalline CVD diamonds. Furthermore we investigate the influence of the implantation parameters and discuss the optical properties of single Ni-related centers with regard to their applicability as single-photon sources.

\section{EXPERIMENTAL}

For our investigations we used type IIa CVD single-crystal plates (Element Six, Isle of Man). They are specified with a nitrogen mass content of less than \SI{1}{ppm} and a boron mass content of less than \SI{0.05}{ppm}.

Ni ions were implanted with energies between \SI{30}{keV} and \SI{1810}{keV} and doses between \SI{2E8}{ions/cm^2} and \SI{E14}{ions/cm^2}. Monte Carlo simulations (SRIM\footnote{Stopping range of ions in matter: http://www.srim.org/}) yield mean Ni implantation depths between \SI{16}{nm} (\SI{30}{keV}) and \SI{793}{nm} (\SI{1810}{keV}). Implantations with energies up to \SI{400}{keV} were performed at the Georg-August Universit\"at (G\"ottingen, Germany), higher energy implantations at the RUBION (Ruhr-Universit\"at, Bochum, Germany). After implantation, the samples were annealed at \SI{1000}{\celsius} in vacuum for one hour followed by \SI{15}{minutes} of cleaning in oxygen plasma to remove graphite from the sample surface formed during the annealing.

To analyze our diamond samples we use a home-built confocal microscope: The fluorescence is excited with a cw \SI{671}{nm} DPSS laser through a microscope objective with $\text{NA} =0.8$. A dichroic mirror and dielectric filters block the excitation light from entering the analyzing beam path and allow only fluorescence light with $\lambda > \SI{720}{nm}$ to pass. The fluorescence can be guided into a grating spectrometer to analyze the fluorescence spectrum. Time correlation of the photoluminescence signal can be measured with a \emph{Hanbury-Brown-Twiss} setup (HBT) consisting of two single-photon sensitive avalanche photodiodes (APDs, Perkin Elmer SPCM-AQR-14) on each side of a $50/50$ beam splitter. Bandpass filters were placed in front of the APDs to prevent crosstalk. A fast counting electronics (PicoQuant PicoHarp 300) records two lists of photon arrival times; by correlation of these lists one gets a histogram of time delays between the arrival of single photons. It can be normalized to Poissonian statistics by the factor $1/(N_1N_2wT)$, where $N_{1,2}$ are the count rates of each APD, $w$ is the time-bin-width of the histogram, $T$ is the complete correlation time\cite{Brouri2000}. After correction of random coincidences due to uncorrelated background emission\cite{Brouri2000} one yields the \emph{second-order correlation function}
\begin{equation}
   g^{(2)}(\tau)=\frac{\langle I(t)I(t+\tau)\rangle}{\langle I(t)\rangle^2}
\end{equation}
of the fluorescence light, representing the probability to detect two photons in the time interval $\tau$. Due to the timing jitter of the APDs there is a \emph{device-response function} (DRF) the measured $g^{(2)}$ has to be corrected with. By recording the $g^{(2)}$-function of \SI{100}{fs} long laser pulses of a \SI{80}{MHz} Ti:Sapphire the combined DRF of our HBT setup was determined as a gaussian peak function with a half $1/\sqrt e$-width of $w=\SI{354}{ps}$. The intensity signal of the APDs can also be used for confocal scans of the diamond sample using a three axis step motor system. Limited by the focus size of the excitation light a point-like emitter appears with a lateral size of $\approx\SI{1.1}{\micro m}$ and an axial size of $\SI{3.5}{\micro m}$ in the scans. The complete detection efficiency $\eta$ of our setup consists of the collection efficiency of the objective, the detection efficiency of the APDs and the transmission of all optical components. Disregarding internal reflections at the sample surface we can estimate $\eta=\SI{2.2}{\%}$. The fluorescence can alternatively be guided into a grating spectrometer to analyze the fluorescence spectrum.

\section{RESULTS AND DISCUSSION}

\subsection{Implantation parameters}

\begin{figure}\centering
   \subfigure[]{\includegraphics[height=5.75cm]{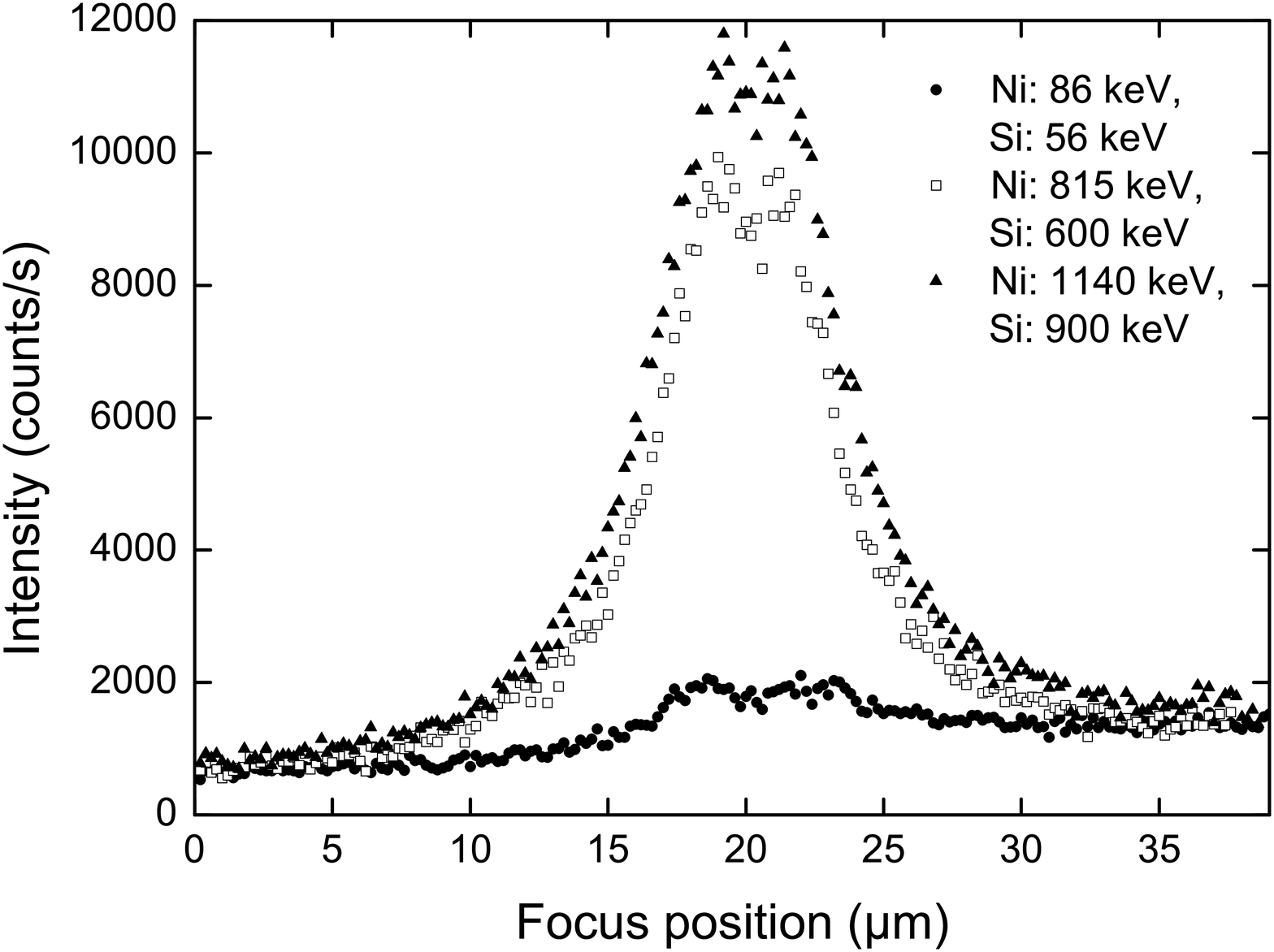}\label{energien}}
   \hspace{1cm}
   \subfigure[]{\includegraphics[height=5.75cm]{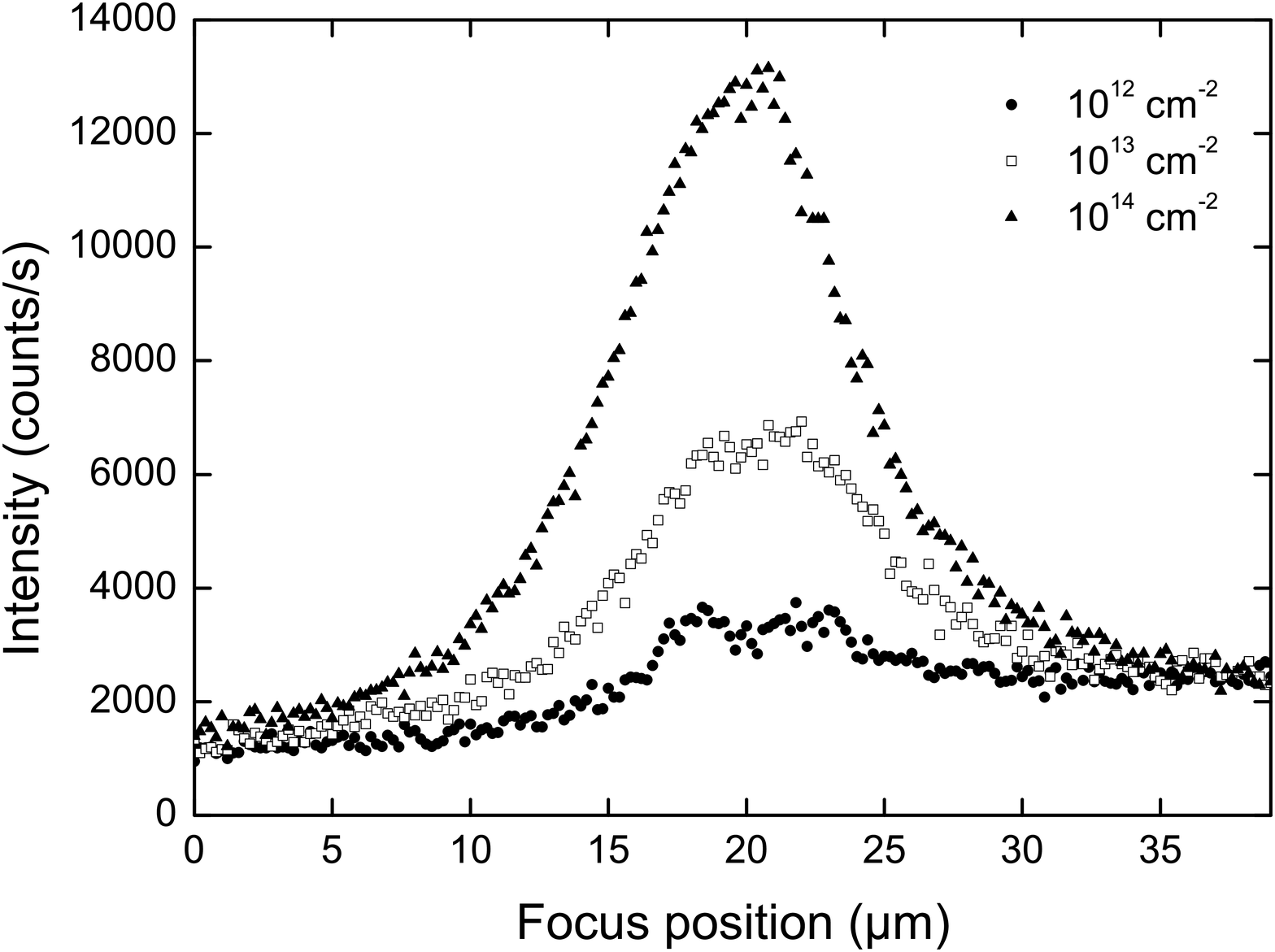}\label{fluenzen}}
   \caption{(a) Depth scan of three samples implanted with Ni and Si at a dose of \SI{E12}{cm^{-2}} and different energies. In the region of the mean implantation depth the high ion radiation damage background in the sample implanted with the higher energies is obvious.\quad(b) Depth scans of three samples implanted with Ni and Si at three different doses. Despite the low implantation energies of \SI{86}{keV}, \SI{56}{keV} respectively the background is clearly visible for the samples with the higher doses.}
\end{figure}
As there is a current tentativeness on the atomic structure of Ni-related defects\cite{Iakoubovskii2002,Yelisseyev2003} and as there have been recent reports on bright color centers based on Si-Ni-complexes\cite{Aharonovich2008,Aharonovich2009}, we tested co-implantation of different combinations of ions: Together with Ni we implanted Si and/or N. The co-implantation of N shall at first support the creation of NE8 centers\cite{Nadolinny1999}, a second aspect concerns a phenomenon observed for negatively charged SiV-centers: Co-implanted N acts like an electron donor leading to a higher Fermi level in the diamond and higher emission rates of the single centers \cite{Wang2008}.

As mentioned above, we used different implantation energies and doses to find the parameters for the optimum color center density, i.\,e. a density that is as large as possible but still allows for the localization of single centers. The variation of implantation energy and the associated different implantation depths reflects a trade-off between two different aspects: If, on the one hand, the distance of a single emitter from the diamond surface is larger than half the emission wavelength, refraction at the surface becomes relevant, lowering the collection efficiency. On the other hand, surface effects might act detrimental on color centers if the distance to the surface is too small. In this regard, NV centers were reported to be present in their negatively charged state (i.\,e. emission at \SI{638}{nm}) only if they are situated in a depth in the sample larger than \SI{200}{nm}\cite{Santori2009}.

Photoluminescence investigations of the implanted samples reveal the emergence of a broadband fluorescence background (measured in the interval \SI{765}{nm}-\SI{815}{nm}) due to radiation damage for too high implantation energies and too high implantation doses. As an example Fig. \ref{energien} shows this background fluorescence between \SI{765}{nm} and \SI{775}{nm} as a function of the focus position for samples implanted with Ni and Si at different energies and an ion dose of \SI{E12}{cm^{-2}}. This dose was found to be the lower limit of ion doses in order to produce a detectable number of color centers. From Fig. \ref{energien} it is clearly visible that high energy implantation (Ni: $\geq$ \SI{815}{keV}, Si: $\geq$ \SI{600}{keV}) produces a considerable background fluorescence in the region of the mean implantation depth. This high background in the spectral range of Ni-center emission prohibits the detection of single emitters. For the lower implantation energy shown here there still emerges some background fluorescence but at a level tolerable for single emitter detection. Fig. \ref{fluenzen} presents measurements of the background fluorescence for a fixed implantation energies (Ni: \SI{86}{keV}, Si: \SI{86}{keV}) and different ion doses. These results reveal that also high doses ($\geq\SI{E13}{cm^{-2}}$) generate unwanted broadband fluorescence. Taking into account these results, we used diamonds implanted with low energy ions (Ni: \SI{86}{keV}, Si: \SI{56}{keV}, N: \SI{30}{keV}) and a dose of \SI{E12}{cm^{-2}} for investigations on single emitters.

\subsection{Spectroscopy}

\begin{figure}\centering
   \subfigure[]{\includegraphics[height=5.5cm]{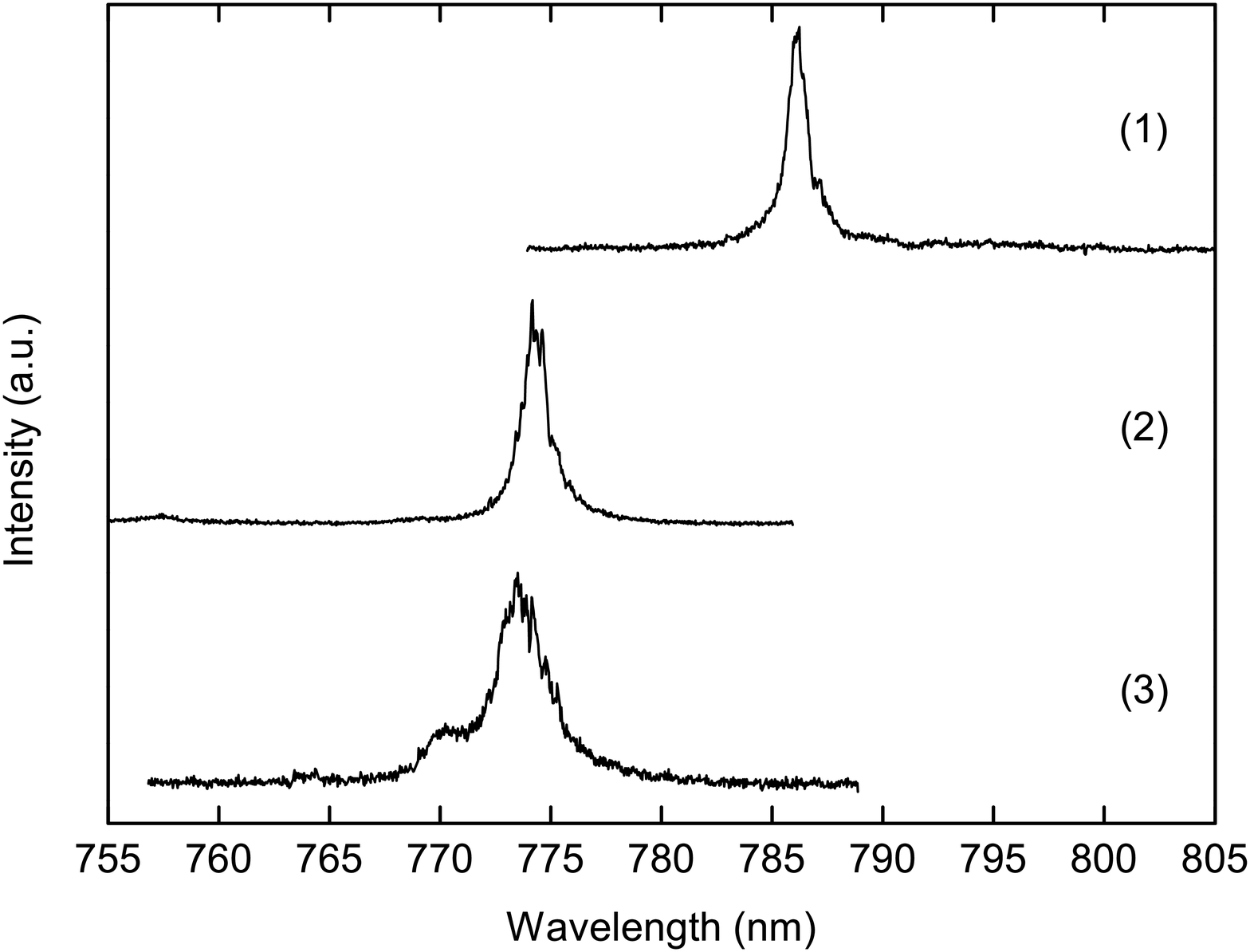}\label{spektren}}
   \hspace{1cm}
   \subfigure[]{\includegraphics[height=6.75cm]{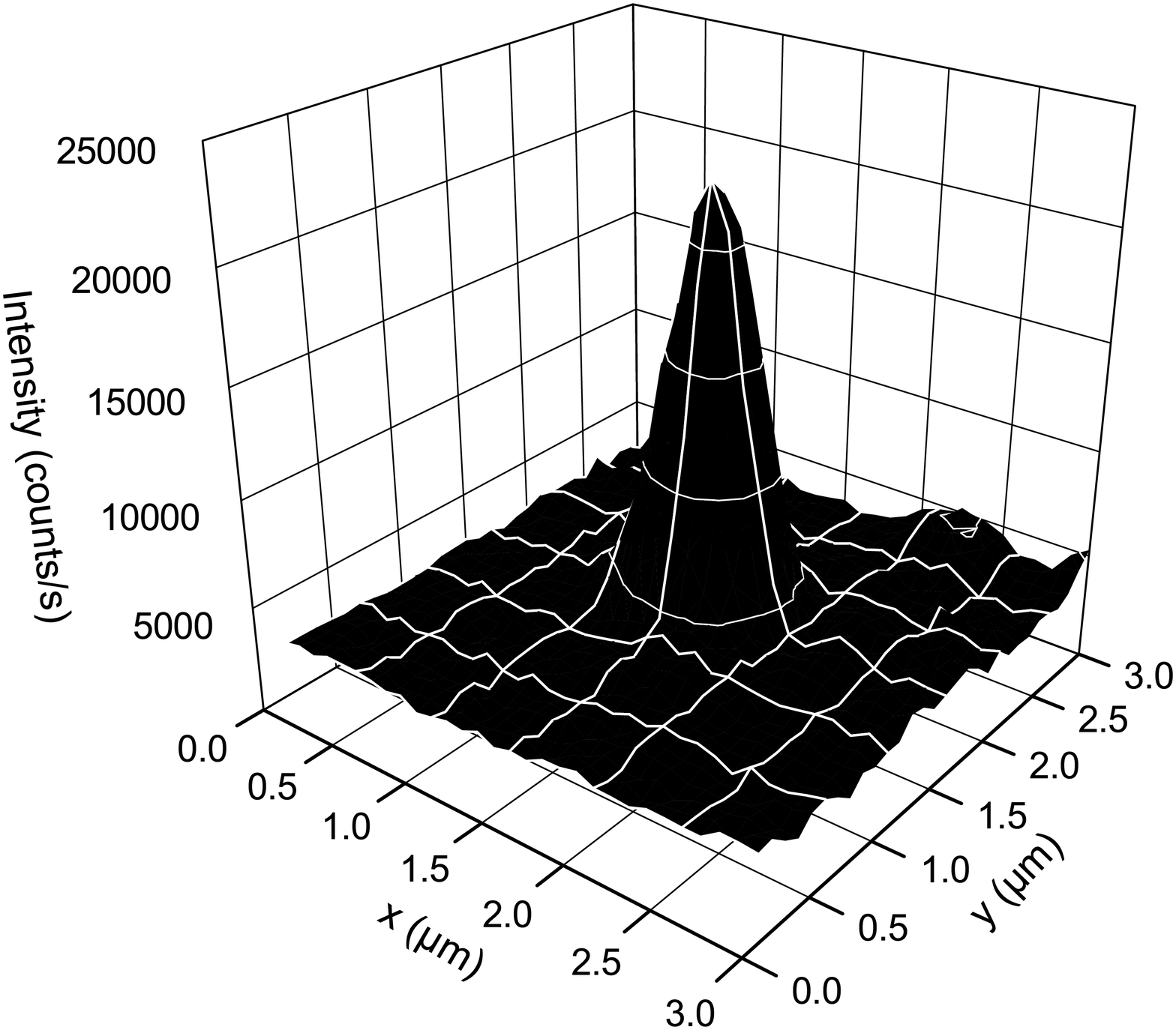}\label{scan}}
   \caption{(a) Fluorescence spectra of three different single Ni-related color centers. The emission wavelengths are detected
      between \SI{767}{nm} and \SI{788}{nm} and the line-widths amount to \SI{2}{nm} to \SI{3}{nm} FWHM.\quad(b) Confocal scan of emitter (3). The size of the emitter in the scan is limited by the objective focus size.}
\end{figure}
In order to investigate the different samples we performed confocal scans, detecting the fluorescence intensity in spectral windows of typical Ni-center emission (\SI{765}{nm}-\SI{815}{nm}). For the samples implanted with Ni only and with Ni and N we could not detect any emission within the detection windows. In samples implanted with Ni and Si, however, it was possible to locate several single color centers even though with a rather low density, i\,e. approximately one center per $\SI{85}{\micro m}\times\SI{85}{\micro m}$-scan. Given the implantation dose of \SI{E12}{cm^{-2}} this would correspond to a formation efficiency of \SI{1.4E-8}{}. The color center emission lines are in the spectral range between \SI{770}{nm} and \SI{786}{nm}. Emission spectra of three centers are presented in Fig. \ref{spektren}.

Fig. \ref{scan} shows a confocal scan of  emitter (3). The size of the emitter in the scan image is limited by the spatial resolution of the confocal microscope. Its fluorescence spectrum (Fig. \ref{spektrum}) can be fitted by two Lorentzian peaks at center wavelengths of \SI{770.0}{nm} and \SI{773.6}{nm} with line-widths of \SI{1.36}{nm} and \SI{2.70}{nm} FWHM. These spectra are similar to those reported for Ni-Si complexes in diamond nano-crystals\cite{Aharonovich2008,Aharonovich2009}. A vibronic sideband could not be observed.

\subsection{Single-photon emission}

\begin{figure}\centering
   \subfigure[]{\includegraphics[height=5.75cm]{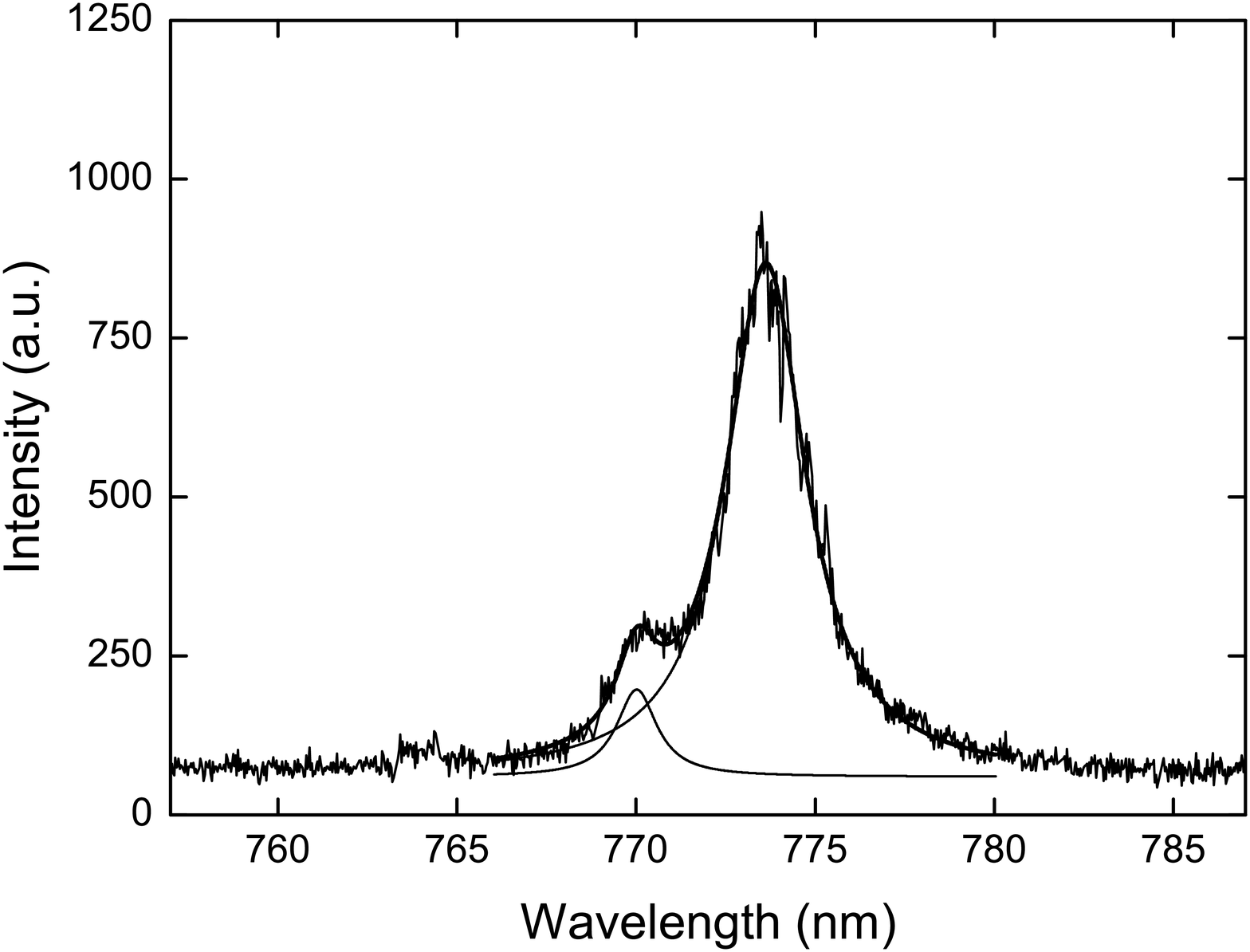}\label{spektrum}}
   \hspace{1cm}
   \subfigure[]{\includegraphics[height=5.75cm]{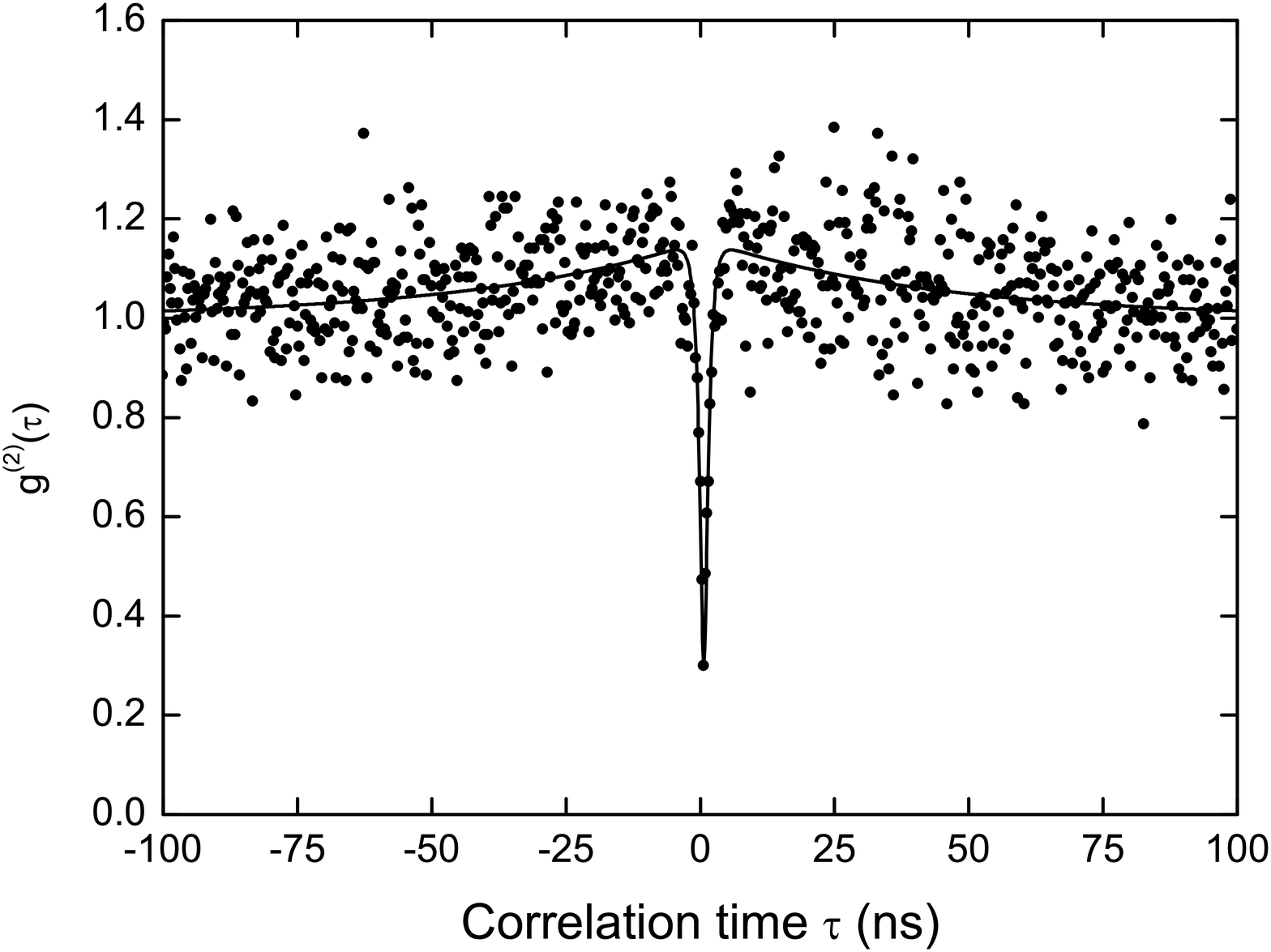}\label{korrelation}}
   \caption{(a) Emission spectrum of emitter (3) of Fig. \ref{scan}. The spectrum can be fitted with a sum of two Lorentzian
      peaks, yielding central wavelengths of \SI{770.0}{nm} and \SI{773.6}{nm} and line-widths of \SI{1.36}{nm} and \SI{2.70}{nm}. A vibronic sideband at larger wavelengths could not be observed.\quad(b) Second-order correlation function single emitter (3). The difference of $g^{(2)}_{\text{meas}}(0)$ from zero is due to the timing jitter of the APDs, namely \SI{354}{ps}. For the fit function a three-level model is used (see text), that is convoluted with the gaussian time response function of the HBT setup. The fit function predicts $g^{(2)}_{\text{fit}}(0)=0.31$, the deviation from the measured $g^{(2)}_{\text{meas}}(0)$ is 0.008.}
\end{figure}
The sub-Poissonian character of the single emitter's emission can be demonstrated by recording the second-order correlation function $g^{(2)}$: In Fig. \ref{korrelation} the normalized $g^{(2)}$-measurement of color center (3) is shown, where the data has been corrected for background events according to the signal-to-noise ratio (SNR) of 6:1, obtained from the confocal scan in Fig. \ref{scan}. During the measurement no blinking or bleaching was observed. The value of $g^{(2)}_{\text{meas}}(0)$ is clearly smaller than 0.5, indicating the presence of a single emitter. Nevertheless, the measurement shows photon bunching: For time scales $\tau >\SI{5}{ns}$, $g^{(2)}$ exceeds the value of one, which can be explained by the existence of a longer lived shelving state (in the equations below signed as state ``3'') besides ground (``1'') and excited state (``2''). In the following, this shelving state is supposed to decay into the ground state only and this transition is assumed to be non-radiative. The extent of the bunching is depending on the two inter-system crossing rates for the transitions between excited state and shelving state and between shelving state and ground state. Analogous to the considerations of Kurtsiefer et al.\cite{Kurtsiefer2000}, an analytical form of the second order correlation function of such a system is given as
\begin{equation}\label{g2-funktion}
   g^{(2)}(\tau)=1-(1+a)\,\text{e}^{-|\tau|/\tau_1}+a\,\text{e}^{-|\tau|/\tau_2}\,.
\end{equation}
If  $r_{ij}$ represents the transition rate from state $i$ to state $j$, the coefficients are
\begin{equation}
   \begin{array}{c}
      \tau_{1,2}=2/(A\pm\sqrt{A^2-4B})\,,\\[0.5em]
      a         =\frac{1-r_{31}\tau_2}{r_{31}(\tau_2-\tau_1)}
   \end{array}
\end{equation}
with the abbreviations
\begin{equation}
   \begin{array}{c}
      A=r_{12}+r_{21}+r_{23}+r_{31}\,,\\[0.25em]
      B=r_{12}r_{23}+r_{12}r_{31}+r_{21}r_{31}\,.
   \end{array}
\end{equation}
Equation \ref{g2-funktion} predicts $g^{(2)}(0)=0$, which deviates from the measured $g^{(2)}_{\text{meas}}(0)=0.30$ of Fig. \ref{korrelation}. The reason is the limited time resolution of the HBT setup due to the timing jitter of the APDs. Thus, the measured $g^{(2)}_{\text{meas}}(\tau)$ of Fig. \ref{korrelation} has to be fitted with a convolution of the ideal $g^{(2)}(\tau)$ (Eqn. \ref{g2-funktion}) with the device-response function:
\begin{equation}\label{fit-funktion}
   g^{(2)}_{fit}(\tau)=\frac{1}{\sqrt{2\pi}w}\int\limits_{-\infty}^{\infty}g^{(2)}(\tau')\cdot\text{e}^{-(\tau -\tau')^2/2w^2}\text{d}\tau'\,.
\end{equation}
Fitting Eqn. \ref{fit-funktion} to the data yields constants of $\tau_1=\SI{0.83}{ns}$, $\tau_2=\SI{42.2}{ns}$ and $a=0.16$. The deviation of $g^{(2)}_{\text{fit}}(0)=0.31$ predicted by the fit function and the measured value is 0.008. This small deviation proves that the difference of the measured $g^{(2)}_{\text{meas}}(0)$ from zero is solely due to the limited timing resolution of our photon counting setup and the emitter shows perfect anti-bunching.

\begin{figure}\centering
   \subfigure[]{
      \includegraphics[height=5.75cm]{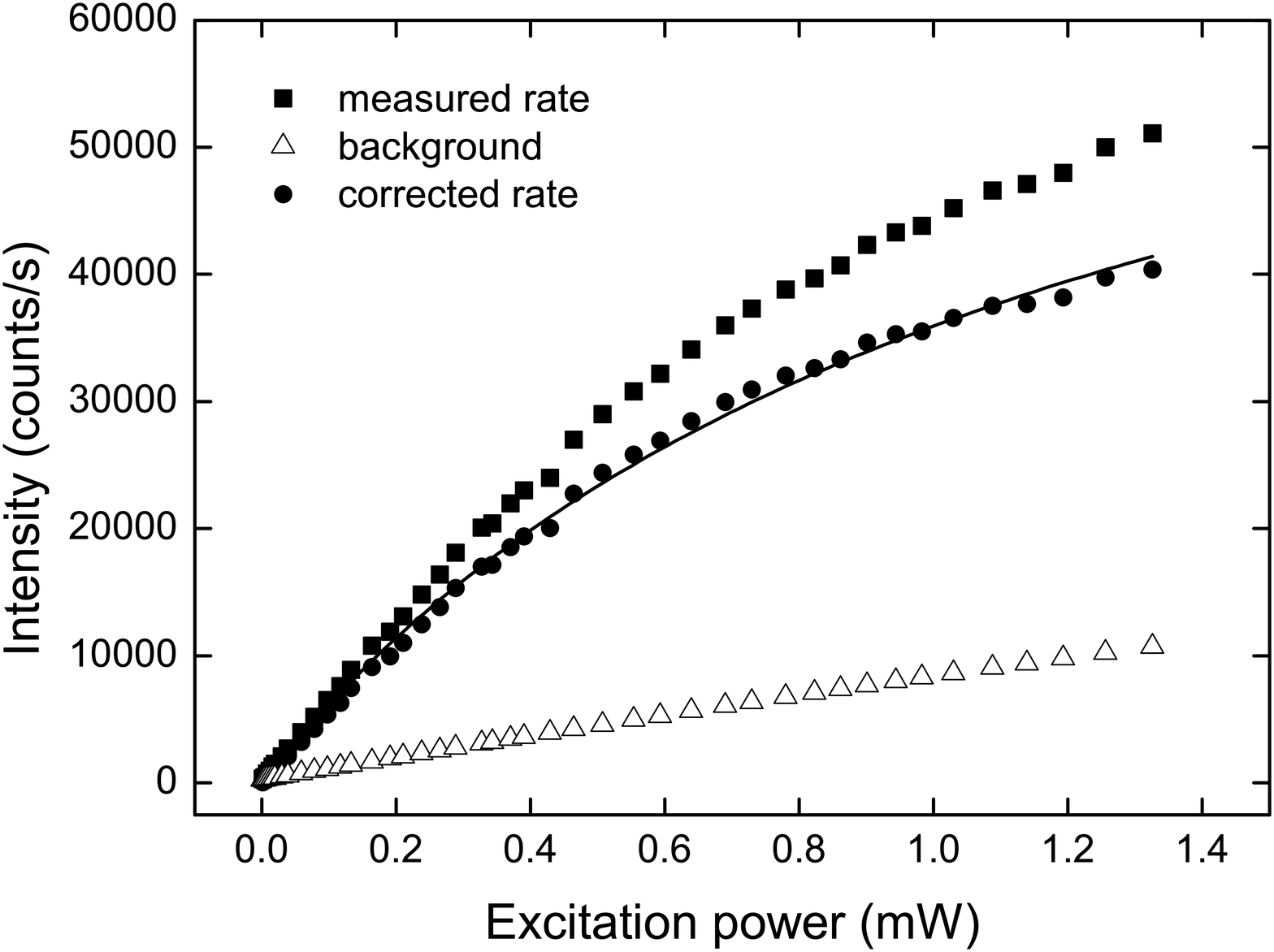}\label{saettigung}}
      \hspace{1cm}
   \subfigure[]{\includegraphics[height=5.75cm]{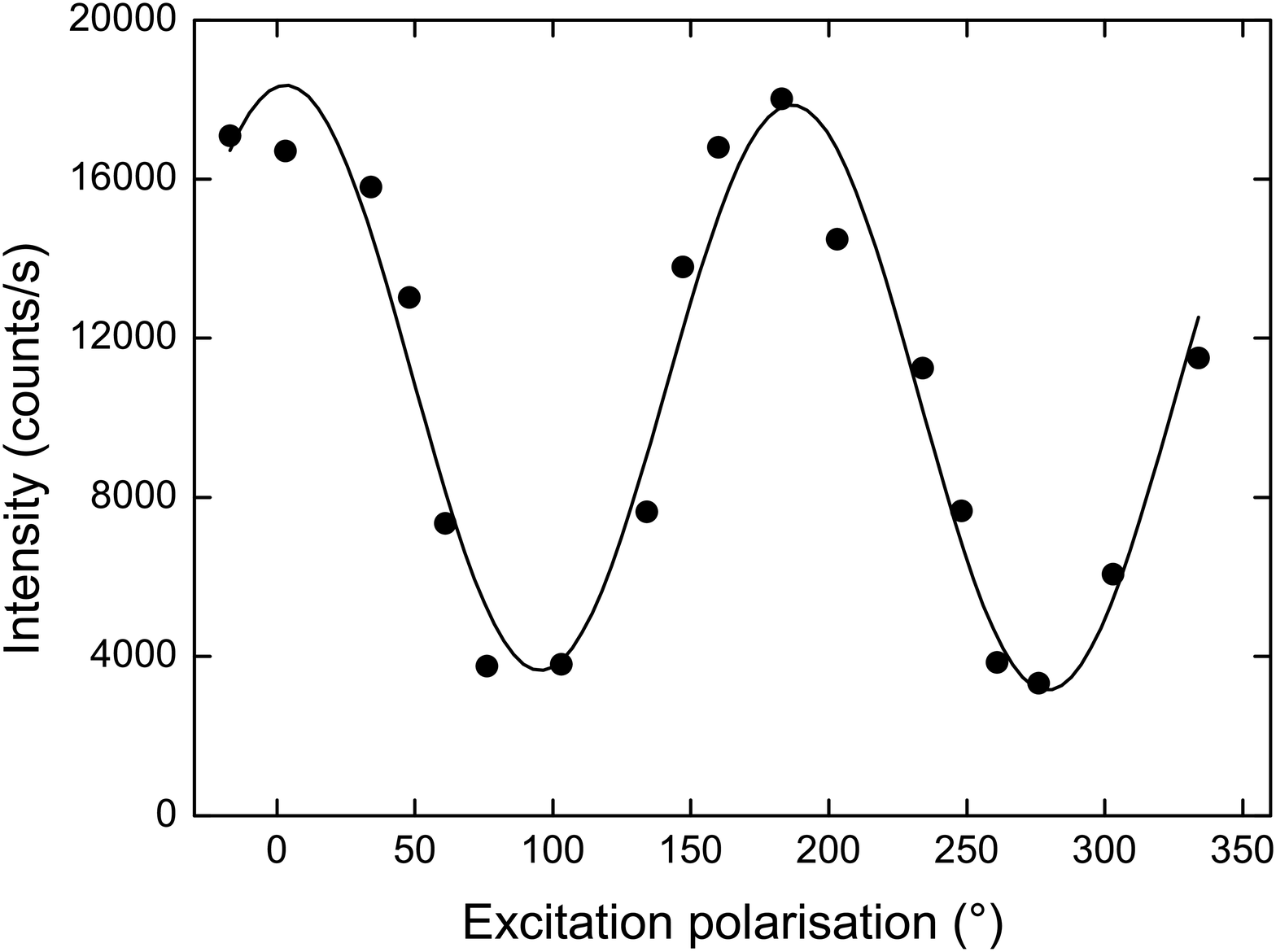}\label{polarisation}}
   \caption{(a) Saturation measurement on color center (3). The measured count rate was corrected with the nearly linear
      background measured \SI{1}{\micro m} beside the emitter. The saturation count rate is \SI{77.9}{kcount/s}, the saturation intensity was determined to be \SI{365}{kW/cm^2}.\quad(b) Intensity as function of the polarization angle of the excitation light. An angle of \ang{0} corresponds to the $\langle 001\rangle$ crystal directions.}
\end{figure}
We also performed a saturation measurement by increasing the excitation power and recording the total count rate of the color center with the HBT setup. The background emission was measured at a spot \SI{1}{\micro m} beside the color center and subtracted from the signal. The result is displayed in Fig. \ref{saettigung}. A fit for the count rate $R$ of the form
\begin{equation}
   R=R_\infty\frac{P}{P+P_\text{sat}}
\end{equation}
to the data of Fig. \ref{saettigung} yields a saturation power $P_\text{sat}$ of \SI{1.17}{mW} and a saturation count rate $R_{\infty}$ of \SI{77.8}{kcount/s}. This value is of the same magnitude as reported for NE8 centers\cite{Wu2006} but about \SI{50}{\%} lower than the \SI{200}{kcounts/s} reported for Ni/Si-centers in nano-diamonds\cite{Aharonovich2009}. With a half $1/\sqrt e$ focus width of \SI{223}{nm}, also estimated from Fig. \ref{scan} one yields a saturation intensity of \SI{365}{kW/cm^2}. For a rough estimate of the radiative excited state lifetime $\tau_f$ of the emitter we assume a two-level system. $\tau_f$ can than be calculated from $\tau_1$ by applying a correction factor taking into account the excitation power (the $g^{(2)}$ of Fig. \ref{korrelation} was recorded at \SI{390}{\micro W}) and the saturation power (\SI{1.17}{mW}, see above)\cite{Trebbia2009}:
\begin{equation}
   \tau_f=\tau_1\cdot\left(1+\frac{P}{P_\text{sat}}\right)=\SI{0.83}{ns}\cdot1.33 =\SI{1.11}{ns}\,.
\end{equation}
This value is comparable to the lifetime of \SI{2}{ns} determined for Ni-related centers\cite{Wu2007,Aharonovich2009}. To measure the radiative excited state lifetime with higher precision one has to determine all four transition rates of equation \ref{g2-funktion} with the help of an excited state lifetime measurement or with excitation power dependent $g^{(2)}$ measurements. This work is in progress.

To gain insight into the emitter's orientation in the diamond crystal, the dependency of the count rate on the polarization angle of the excitation light was measured (Fig. \ref{polarisation}). It can be well described by a sinusoidal variation. From the positions of the intensity maxima and the known orientation of the fcc diamond lattice in our setup, one can deduce the alignment of a dipole or at least its projection to the $\langle 001\rangle$ direction. However, the visibility of the polarization-dependent excitation,
\begin{equation}
   V=\frac{I_{max}-I_{min}}{I_{max}+I_{min}}\,,
\end{equation}
amounts to \SI{65}{\%} only. The fact that this value is smaller than \SI{100}{\%} might be attributed to the presence of a second dipole at an orthogonal direction. A similar behavior is known from the NV center for which electric dipole transitions are allowed for dipoles in a plane perpendicular to the NV symmetry axis\cite{Davies1976} or for the \SI{1.40}{eV} Ni-related center where a differently polarized zero-phonon-line doublet arises due to two possible dipole orientations in the diamond lattice\cite{Collins1989}. Further investigations are required to explore the dipole structure of the Ni-Si complexes in more detail.

Finally, we want to discuss the atomic composition of the presented color centers. The spectral range of the emission wavelengths of our color centers partly overlaps with the range from \SI{782}{nm} to \SI{801}{nm} reported for NE8 centers\cite{Wu2006,Wu2007,Rabeau2005,Gaebel2004}. However, the samples that have been co-implanted with Ni and N do not show emission of color centers, whereas the centers presented here have been detected after co-implantation of Ni and Si. We therefore suppose that the presented centers are complexes containing Ni and Si. As far as the involvement of N is concerned, no definite statement is possible because of the N fraction present before implantation. We hope to answer this question with implantations in diamond containing less N.

\subsection{Conclusion}

We report observation of the first single Ni-related color centers in single crystalline diamond produced via ion implantation, which is an important step towards their deterministic production. With an implantation depth of $\approx\SI{40}{nm}$ they are close to the surface and exhibit emission lines between \SI{770}{nm} and \SI{786}{nm} with line-widths of about \SI{3}{nm} FWHM. A measured $g^{(2)}(0)$ of 0.31 can completely be explained with the timing resolution of the HBT setup and the very short lifetime of the excited state and unambiguously proves single-photon emission. The saturation count rate was determined as \SI{77}{kcounts/s}. Polarization measurements indicate that two orthogonal dipoles contribute to the fluorescence. Different combinations of implantated ions lead to the conclusion of Si being constituent of the centers. With further variation of implantation parameters and co-dopants we aim at enhancing the creation efficiency and having more detailed information about the atomic structure.

\section{Acknowledgements}

The ion implantations were performed with Prof.\,Dr.\,Wolfgang Bolse (Universit\"at Stuttgart, Germany) at the Universit\"at G\"ottin\-gen, Germany and with PD\,Dr.\,Jan Meijer and Dr.\,Detlef Rogalla at the RUBION (Ruhr-Universit\"at Bochum, Germany). The project was financially supported by the Deutsche Forschungsgemeinschaft and the Bundesministerium f\"ur Bildung und Forschung (EphQuaM network, contract 01BL0903).


\end{document}